# Machine learning driven high-resolution Raman spectral generation for accurate molecular feature recognition

Vikas Yadav,[‡] Abhay Kumar Tiwari,[⊥] Soumik Siddhanta,[‡*]

[‡]*Department of Chemistry, Indian Institute of Technology Delhi, Hauz Khas, New Delhi, India, 110016*

[⊥]*SPATIALTY.AI, Tagya Research Private Limited, Rajankunte, Bangalore 560064*

*Corresponding author e-mail: soumik@iitd.ac.in

**Abstract**

Through the probing of light-matter interactions, Raman spectroscopy provides invaluable insights into the composition, structure, and dynamics of materials, and obtaining such data from portable and cheap instruments is of immense practical relevance. Here, we propose the integration of a Generative Adversarial Network (GAN) with low-resolution Raman spectroscopy with a portable hand-held spectrometer to facilitate concurrent spectral analysis and compound classification. Portable spectrometers generally have a lower resolution, and the Raman signal is usually buried under the background noise. The GAN-based model could not only generate high-resolution data but also reduced the spectral noise significantly. The generated data was used further to train an Artificial Neural Network (ANN)-based model for the classification of organic and pharmaceutical drug molecules. The high-resolution generated Raman data was subsequently used for spectral barcoding for identification of the pharmaceutical drugs. GAN also demonstrated enhanced robustness in extracting weak signals compared to conventional noise removal methods. This integrated system holds the potential for achieving accurate and real-time monitoring of noisy inputs to obtain high throughput output, thereby opening new avenues for applications in different domains. This synergy between spectroscopy and machine learning (ML) facilitates improved data processing, noise reduction, and feature extraction and opens avenues for predictive modeling and automated decision-making using cost-effective portable devices.

# Introduction

Raman spectroscopy, a non-destructive optical tool for high-sensitivity chemical analysis of molecular structures, molecular interactions, phase, polymorphism, crystallinity, and effects of light-matter interactions, has witnessed intense growth over the years, finding applications across diverse scientific domains.[1,2,3] The inelastic scattering of light, when interacting with molecules or crystals, generates higher (anti-stokes) and lower energy photons (stokes). These shifts of frequencies correspond to the vibrations present in the molecules, and the interpretation of such vibrations provides exquisite molecular structural information, making it an invaluable tool for elucidating the intricacies of chemical structures and offering insights into molecular conformations, compositions, and interactions.[4] Despite the promise of widespread use in materials science, pharmaceutical industries, and medicine, the widespread adoption of Raman spectroscopy for everyday applications has been hampered by the requirement of high-end spectrometers with low inherent noise and high sensitivity.[5,6,7,8]

There are several ways to enhance Raman signal intensities and resolution through hardware and software interventions.[9,10] However, for the widespread implementation of Raman spectroscopy, portable instruments are required which are safe to use and cost-effective. Some spectrometers are portable and dispersive, but there is a fundamental limit to light collection, which necessitates the use of cooled detectors and high-power lasers, driving the cost up significantly. There have been reports of miniature chip-scale low-power laser excitations and swept-source spectrometers being used to capture the Raman scattering, where one in a million to a billion photons gets scattered and needs to be detected.[11] The diffusion and scattering of photons from inhomogeneous samples also create additional complications due to the need to have high collection efficiency. The scattered light is collected using an optical system consisting of lenses, filters, and optical fibers. It is directed towards a dispersive spectrometer and a detector, either a charge-coupled device (CCD) or complementary metal-oxide-semiconductor (CMOS), where the optical input is converted into an electric signal whose strength is proportional to the total photon flux. The excitation laser can also lead to fluorescence, which the detector can pick up and restrict the Raman analysis to strong bands only. Detectable signals for identifying chemical characteristics can, therefore, be hindered by noises and interferences, which, along with fluorescence background, cannot be removed by longer accumulation times or using higher laser powers. Several denoising methods are used to increase the signal-to-noise ratio (SNR) and remove shot noise and fluorescence backgrounds, such as discrete wavelet transform (DWT) and adaptive lifting wavelet transform (ALWT).[12,13,14] Other methods, such as frequency filtering using the Savitzky-Golay algorithm, can also be used for post-processing and noise removal from the spectrum.[15] The background noise can also arise due to laser fluctuations, pixel-to-pixel variations, and dark current, which is dependent on the temperature and its fluctuations, necessitating the reduction of the detector temperature.[16] Additionally, the readout noise creates inaccuracies in the photon-to-voltage conversion and, therefore, requires amplifying the generated signal and sampling it by an analog-to-digital converter.[17] The noise reduction is also achieved using a lock-in amplifier, where the frequency component with similar modulation of the incident laser is filtered out, improving the SNR.[18] The background noise can also be rejected by tuning the excitation wavelength or diffraction grating or spatially offset Raman.[10, 19] Other methods, such as surface-enhanced Raman scattering (SERS), can also be employed to increase the Raman signal and to enhance the SNR.[20,21] Also, different Raman spectrometers can have different instrument responses, thus giving variable intensities and spectral shifts for similar molecular entities.[22] All these factors can significantly affect quantitative analysis. As an alternative approach, we propose a machine learning-based technique to convert the low-resolution Raman spectrum into a high-resolution spectrum, making it possible to use any spectrometer, typically used for absorbance and fluorescence measurements, for sensitive Raman measurements.

Integrating machine learning (ML) with Raman spectroscopy marks a convergence of analytical chemistry and artificial intelligence, promising to revolutionize how we extract information from complex spectral data.[23,24,25] Raman spectroscopy, while powerful in capturing molecular fingerprints, often faces challenges in dealing with inherently noisy and low-resolution spectra, mainly when acquired from miniaturized spectrophotometers. A few methods, such as direct standardization and piecewise direct standardization, can map data from one instrument to another by calculating a transfer matrix between the source and the target instruments.[26,27] A method of spectral recognition analogous to facial recognition has also been used to match the SERS spectrum to standard Raman libraries using an ML algorithm using a special metric of "Characteristic Peak Similarity".[21] This method tries to mitigate the effects of substrate differences in SERS measurements and map and identify them accurately. Chemometric methods, in general, usually work well in linear mapping but are not useful in non-linear fluctuations, which is seen in various instruments with differences in hardware components.[28] The non-linearity of the data also presents a significant challenge in adopting mathematical functions for spectral analysis. Spectroscopic classification and Transfer-learning-based Raman spectra identification have been attempted using a deep convolution network (CNN), which performs better than non-deep ML algorithms such as support vector machines (SVM) in real-world multiclass classification tasks.[29,30] Transfer learning further reduces the need for large data training sets through the adaptation of the learned knowledge from the source domain to the target domain.[31] However, all known methods can identify or map spectra obtained only from spectrometers of relatively high resolution. Our research seeks to utilize ML, specifically Generative Adversarial Network (GAN), to bridge the resolution gap, unlock the full analytical capabilities of miniaturized, and enable the detection of Raman scattered photons. GAN has emerged over the past few years and has found significant use in applications such as image generation, spectral transfer, and natural language processing.[22,32] The core of our methodology lies in the utilization of the GAN model to enhance the resolution of low-quality Raman scattering obtained from low-resolution absorption or fluorescence spectrophotometers whose resolution was an order lower than high-resolution Raman spectrometers. The GAN is an advanced ML architecture renowned for its prowess in image-to-image translation tasks.[32] In our case, this involves training the model on a meticulously curated dataset comprising pairs of low-resolution and high-resolution Raman spectra. The GAN model learns the intricate mapping between the two sets, enabling it to generate high-resolution Raman spectra when provided with their low-resolution counterparts. The training process involves a dual-network structure – a generator network responsible for creating high-resolution spectra and a discriminator network trained to discern the authenticity of the generated spectra compared to true high-resolution counterparts. Through an iterative adversarial process, the generator refines its ability to create high-fidelity spectra while the discriminator evolves to become increasingly discerning. The equilibrium between these networks signifies the model's proficiency in producing realistic high-resolution Raman spectra from low-resolution inputs.

Our study holds significant promise for developing portable spectrophotometry devices or bypassing the capabilities of high-cost and high-resolution spectrometers for various applications. Herein, we have a) generated high-resolution, high SNR data from a low-resolution hand-held optical spectrometer, b) employed an Artificial Neural Network (ANN) based training algorithm to test the classification efficiency of the generated set of high-resolution spectra, and c) employed a bar-code approach to classify and identify organic compounds which are pharmaceutically important. The democratization of high-resolution spectroscopic techniques through portable devices holds the potential to transform fields where real-time molecular analysis is imperative, offering a versatile and accessible solution for diverse applications. The seamless integration of the ML algorithm, GAN, with Raman spectroscopy mitigates the trade-off between size and resolution, propelling the fabrication of miniaturized, cost-effective Raman spectrophotometers with a vast high-resolution Raman spectral library already available.

# Experimental Section

## Materials and Methods

4-Nitrothiophenol (4-NTP, 80%) and 4-Mercaptobenzoic acid (4-MBA, 99%) were purchased from Sigma-Aldrich and used without further modification. Ranitidine-150 (RANTAC), N-(2-[(5-[(Dimethylamino)methyl]furan-2-yl)methylthio]ethyl)-N'-methyl-2-nitroethene-1,1-diamine, 2-Acetoxybenzoic acid (Aspirin, 300mg), Ibuprofen (Ibuwell-400), (±)-2-(p-isobutylphenyl)propionic acid, and 4-hydroxyacetanilide (PCM, 500mg), were purchased from the local pharmacies in India.

## Raman Spectral Acquisition

The low-resolution Raman spectra of different samples were collected using the Goyalab Spectrophotometer (LR Spectrometer) with a 785 nm laser having an acquisition time of 15 s and 3 accumulations. The LR spectrometer possesses an approximate resolution of 31 cm$^{-1}$, featuring a grating with 150 lines per millimeter and incorporating a linear CCD array. The collection optics involved an Integrated Photonics System optical fiber probe coupled to a 785 nm laser (model: I0785SP100-T040S). The Raman probe has a central fiber for light input and is surrounded by six fiber endings to collect the scattered light. A ball lens is mounted to focus the light into the sample and collect the reflected light. The focal length of this probe is 4 mm. Raman spectra were acquired from aspirin, ibuprofen, paracetamol, ranitidine, 4-MBA, and 4-NTP. Approximately 400 Raman spectra were collected from each sample at different positions to account for spectral variability. Another set of high-resolution spectra were collected using the high-resolution Andor spectrophotometer (HR spectrometer) (Kymera 328i, with quad turret, 328 mm focal length, F/4.1 aperture having CCD camera with resolution < 1 cm$^{-1}$, grating 1800 lines/mm) with similar acquisition setting, i.e., using a 785 nm laser with acquisition time of 15 s and 3 accumulations, and 1800 lines/mm grating system. We also accumulated 400 Raman spectra from each sample using the HR spectrometer. A series of spectra were recorded from each sample to create a comprehensive dataset for training the GAN. Care was taken to minimize external interferences, and all measurements were conducted in a controlled environment to maintain consistency. We have accumulated the spectrum from 200 nm to 1650 nm with 862 wavelength points in between. All the spectra were first normalized and used as input for the ML model.

## Model building for GAN

The GAN model employed in this study consisted of a generator network built upon the U-Net architecture with eight blocks, each comprising convolutional layers, instance normalization, and LeakyReLU activation functions.[32] Low-resolution spectra collected with an integration time of 15 s using an LR spectrometer were used as the input data for the generator. The generator is a combination of an encoder and a decoder structure containing skip connection layers. The generator is crucial in transforming low-resolution Raman spectra into their high-resolution counterparts.[33] The discriminator network consisting of a five-block network which was used to distinguish between generated and true high-resolution spectra. It featured transposed convolution 2D layers, normalization layers, and LeakyReLU activation functions. Training parameters such as the learning rate, batch size, and epochs were carefully selected to optimize model performance. The adversarial training process involved a competitive interplay between the generator and discriminator, with the former generating spectra to deceive the latter and the discriminator evolving to discern true high-resolution spectra. The rigorous evaluation included generalization of performance on unseen spectra, quantitative metrics like SNR, spectral resolution, and peak intensity, and comparative analyses against spectra obtained directly from a high-resolution Raman spectrophotometer.

The loss functions in the GAN model come from both generator ($L_G$) and discriminator ($L_D$) and were defined as[33]

$$L_G = \frac{1}{N} \sum_{n=1}^{N} (D(G(s^n), s^n) - 1)^2 + \lambda \left\| G(s^n) - x^n \right\|_1$$

$$L_D = \frac{1}{N} \sum_{n=1}^{N} (D(x^n, s^n) - 1)^2 + \left( D(G(s^n), s^n) \right)^2$$

Where a high-resolution spectrum is denoted as $x^n$ and a corresponding low-resolution spectrum is denoted as $s^n$, N is the number of training spectra. The hyperparameter λ was introduced to control the magnitude of the L1 norm, a metric used to quantify the dissimilarity between vectors.[33] L1 norm was chosen to improve the SNR of the processed spectra. To train our model, we employed the Adam optimization algorithm over 177 epochs, with a learning rate of 0.0001 and an effective batch size of one.[34] These parameters were selected to optimize the model's performance in reconstructing high-resolution spectra from their low-resolution counterparts.

Classification of GAN-generated spectra using Artificial Neural Network (ANN)

To classify the generated data, we have employed an ANN architecture comprising four layers: an input layer, two hidden layers with 31 neurons each, and an output classification layer with six nodes corresponding to the categories of the samples. The ANN model in this study utilized the Scaled Conjugate Gradient backpropagation algorithm for training, which iteratively updates the weights and biases based on the scaled conjugate gradient (SCG) method.[35] Specifically, we employed the 'trainscg' function and assessed network performance using cross-entropy error and misclassification error metrics.[36,37] During the training state of the ANN model, we set the minimum performance gradient to 1 x e-6 and the maximum validation failure to 6. This means that if the performance of the model does not change by a margin of 1 x e-6 for six successive epochs, training will stop immediately. Upon satisfying these convergence criteria, the model is considered fully trained and ready for testing and validation. To ensure the reliability of our classification model, we conducted a 3-fold validation process by partitioning the dataset into training, testing, and validation subsets. The input data was normalized, and no additional preprocessing steps were applied to promote the generalization of the model. Since we have a substantial dataset, we utilized the entire Raman spectrum without augmentation for both training and testing. High-resolution data was used to train the model. Once the model was trained, it was further tested on the generated high-resolution dataset to assess its performance. The model optimization was based on minimizing mean square error (MSE) and maximizing the coefficient of correlation (R). Convergence criteria for the model were determined by monitoring changes in MSE and R over successive iterations, with the training concluding once a predefined threshold was reached and remained unchanged for a specified number of iterations. The results were further analyzed through visualization techniques, such as confusion matrices and receiver operating characteristic (ROC) curves. Lastly, MATLAB scripts were generated to replicate the results and enable customization of the training process, ensuring reproducibility and flexibility in future experiments.

The MSE for the classification model was calculated as follows:

$$MSE = \frac{1}{n} \sum_{i=1}^{n} (X_i - \widehat{X_i})$$

Wherein n is the number of total observations, $X_i$ is the true class of the sample, and $\widehat{X_i}$ is the class of the sample predicted by the model.

Beyond classification accuracy, efforts were made for post-training interpretability, employing feature analysis and visualization techniques to gain insights into the distinctive spectral attributes defining different sample categories. For the visualization of different categories of samples, we first used Principal Component Analysis (PCA) to reduce the data from higher to lower dimensions. After that, the scores obtained from the PCA were directly incorporated into the Radial Visualization (RadViz) in Orange software.

## Classification of the generated spectra of organic compounds via spectral barcoding

To enable compound classification and automated identification, we adopted the concept of Raman barcoding for the analyte molecules. A comprehensive Raman spectral barcode library was initially constructed using the experimentally obtained Raman spectra from the HR spectrometer. The high-resolution Raman spectra obtained were processed in such a way that all information regarding peak position and full-width half maxima (FWHM) was integrated into the form of barcodes. The thickness and position of vertical lines are directly related to the spectral peak position and FWHM, respectively. An in-house MATLAB script was developed to preprocess the high-resolution spectra into their respective barcodes. Initially, spectral normalization was performed within the intensity range of 0 to 1. Subsequently, a peak searching algorithm identifies peaks in the Raman spectrum based on peak prominence using the "findpeaks" function. Raman barcodes were employed to verify the identity of the sample through barcode comparison. We opted for the Structural Similarity Index (SSIM) method for measuring the similarity between two barcodes. The significance of SSIM lies in its ability to provide a more accurate and human-perceptual measure of image similarity, making it particularly useful in fields such as image processing, computer vision, and remote sensing. SSIM provides a robust and perceptually relevant measure of image similarity, leveraging the structural information within images to offer a meaningful comparison metric. SSIM compares two images, $I$ and $I'$, by considering three components: luminance (*l*), contrast (*c*), and structure (*s*). The SSIM index is calculated using the following formula:[38]

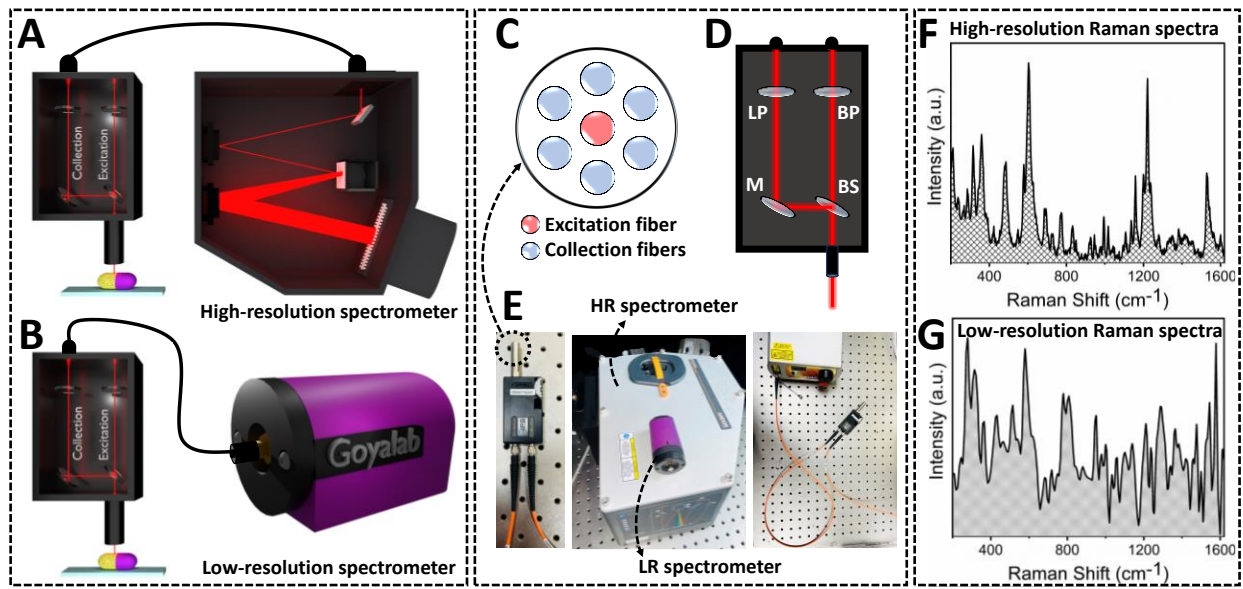

**Figure 1:** (A, B) Shows the schematic representation of the high- and low-resolution Raman spectrometer, respectively. (C, D) Fiber couples collection optics, where the center of the fiber optics is for excitation and the surroundings are for collection of the scattered signal from the sample. The labelled components are, long pass filter (LP), band pass filter (BP), mirror (M), and beam splitter (BS). (E) shows the instrument setup and the perspective views of the low- and high-resolution spectrometers. From left to right are the collection optics, the LR and HR spectrometers and the collection optics connected to the 785 nm laser source. (F, G) shows the Raman spectra of ranitidine drug sample using high- and low- resolution Raman spectrometer, respectively.

$$\text{SSIM}(I, I') = [l(I, I')]^\alpha \cdot [c(I, I')]^\beta \cdot [s(I, I')]^\gamma$$

If a significant portion of the barcode signature of the unknown sample matched with the barcode spectrum from the library, i.e., X, then the unknown sample was identified as X. The threshold percentage of the unknown sample barcode spectrum required for identification is referred to as the percentage match criterion. Subsequently, low-resolution Raman spectra were fed into a GAN model to generate high-resolution spectral outputs. These generated spectra were then encoded into Raman barcodes and compared against the spectral library to determine the similarity, quantified as a percentage match between the unknown sample's Raman barcode and those in the library. Upon scanning the entire library, the model identified the top-scoring profiles, thereby providing the identity of the unknown sample.

## Results and discussion

This study introduces the application of the deep learning technique GAN to effectively transform low-resolution raw Raman spectra, characterized by noise obtained from an LR spectrometer, into high-resolution spectra with substantially improved SNRs. The generated high-resolution spectra were then used to train an ANN model to check the accuracy of compound classification. Spectral barcoding was then performed for efficient compound identification.

### Collection and distinction of high- and low-resolution Raman spectra

Low-resolution Raman spectra were acquired from four different pharmaceutical tablets and two commonly used Raman labels, 4-NTP and 4-MBA, as detailed in the material and method section, using the portable spectrophotometer equipped with a 785 nm laser **(Figure 1)**. The entire experimental setup, including the laser module connected via a fiber optics cable, is compact and portable, aligning with our objective to develop a cost-effective and clinically applicable Raman spectroscopy solution. While advantageous for miniaturization, the LR spectrometer introduces challenges such as a low SNR, causing the signal to be obscured by noise. This limitation underscores the drawbacks associated with downsizing high-resolution Raman spectrometers. The lower grating and smaller focal length of the low-resolution instrument inherently compromise its ability to resolve Raman features effectively. Despite the resolution constraints, our study aims to explore the utility of low-resolution Raman spectrometers compared to their high-resolution counterpart. The high- and low-resolution average Raman spectra for various samples were illustrated in **Figures 2A** and **2B** respectively.

To capture spectral variations comprehensively, approximately 400 spectra were collected for each sample and accumulated at different positions on the samples. The standard deviation in the average spectra plot readily highlights the spectral variability of the system. Intensity box plots were plotted to provide a more adequate representation of this variation, as depicted in **Figure 2E**. The boxplots visually conveyed the spread and comparative distribution of values within a spectral dataset, including key statistical parameters, central tendency, spread, and potential outliers. Given the absence of identifiable spectral features in the noisy low-resolution spectra across various samples, distinguishing between spectra originating from different samples becomes challenging. To enhance the classification performance among distinct sample classes, we employed PCA to reduce the dimensionality of the dataset. PCA is a statistical technique aimed at reducing the dimensionality of datasets while retaining crucial information. PCA yielded score plots, which were subsequently utilized to generate RadViz plots.[39,40] The RadViz plot for the high- and low-resolution Raman data is presented in **Figure 2C** and **2D**, with each dot representing an individual Raman spectrum. The dispersion of points within a specific class in the RadViz plot signifies the extent of spectral variation in the Raman spectra of that particular class. Conversely, a reduced spread in points within a class corresponds to a lower spectral variation. Interestingly, the LR spectral dataset still shows some classification and indicates that the noisy spectra still carry some meaningful information.

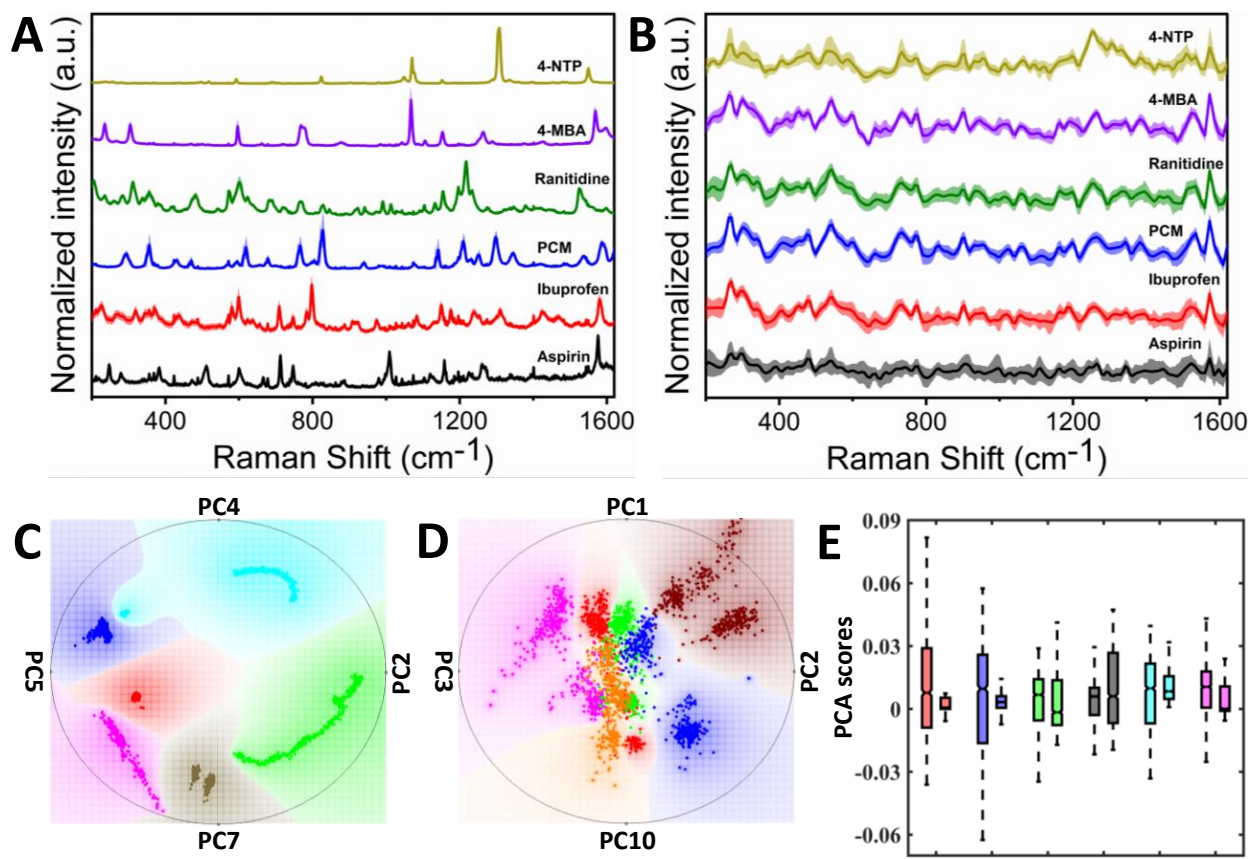

**Figure 2:** Average Raman spectra of different samples accumulated using (A) high resolution and (B) low resolution spectrophotometer. The Radviz plot of the different class of samples using (C) high (D) low-resolution spectrometer. The greater spread represents the more fluctuation in the spectra. (E) Box plot of the different samples using low (left) and high (right) spectrometer. Aspirin (red), ibuprofen (blue), PCM (green), ranitidine (grey), 4-MBA (cyan), and 4-NTP (magenta).

Subsequently, experimental high-resolution Raman spectra were obtained from the same samples through the HR spectrometer, maintaining similar conditions. **Figure 2A** illustrates the high-resolution Raman spectra for various samples, showcasing variations through standard deviation. Evidently, the high-resolution spectra exhibit a greater number of distinguishable features compared to their low-resolution counterparts. The spectral variability in high-resolution spectra was effectively captured using boxplots, depicted in **Figure 2E**. RadViz plots in **Figure 2C** show a better clustering of the high-resolution data.

The Raman band assignment is the most crucial task for identifying a sample and distinct spectral feature for different compounds, which are assigned based on different vibrational bands. All the compounds have Raman bands at different wavenumbers like 643, 791, 850, 1163, 1231, 1273, 1319, 1367, 1558, 1607, and 1644 cm$^{-1}$ mentioned in **Table S1**. The distinct Raman bands observed at 791, 858, and 1231 cm$^{-1}$ correspond to the stretching of the CNC ring, the breathing of the ring, and the stretching of the C-C ring, respectively.[41] All these spectral features are not part of the low-resolution spectra but can be seen clearly in the GAN-generated high-resolution

Raman spectra. The spectral regions around 1300-1350 cm$^{-1}$ correspond to the stretching vibrations of the carboxylate group (-COO$^{-}$).[42] The region between 1580-1610 cm$^{-1}$ contains stretching vibration of the aromatic C=C bond in the phenyl ring.[43] The spectral region around 800-900 cm$^{-1}$ corresponds to the bending vibrations of the C-H bonds in the aromatic ring. Moreover, 1000-1100 cm$^{-1}$ peak is frequently assigned to the stretching vibrations of the C-C and C-O bonds in the aromatic ring.[44,45] In all the compounds where spectra were generated through GAN, the fingerprint region between 200-1800 cm$^{-1}$ is used to differentiate, classify, or identify the molecules.

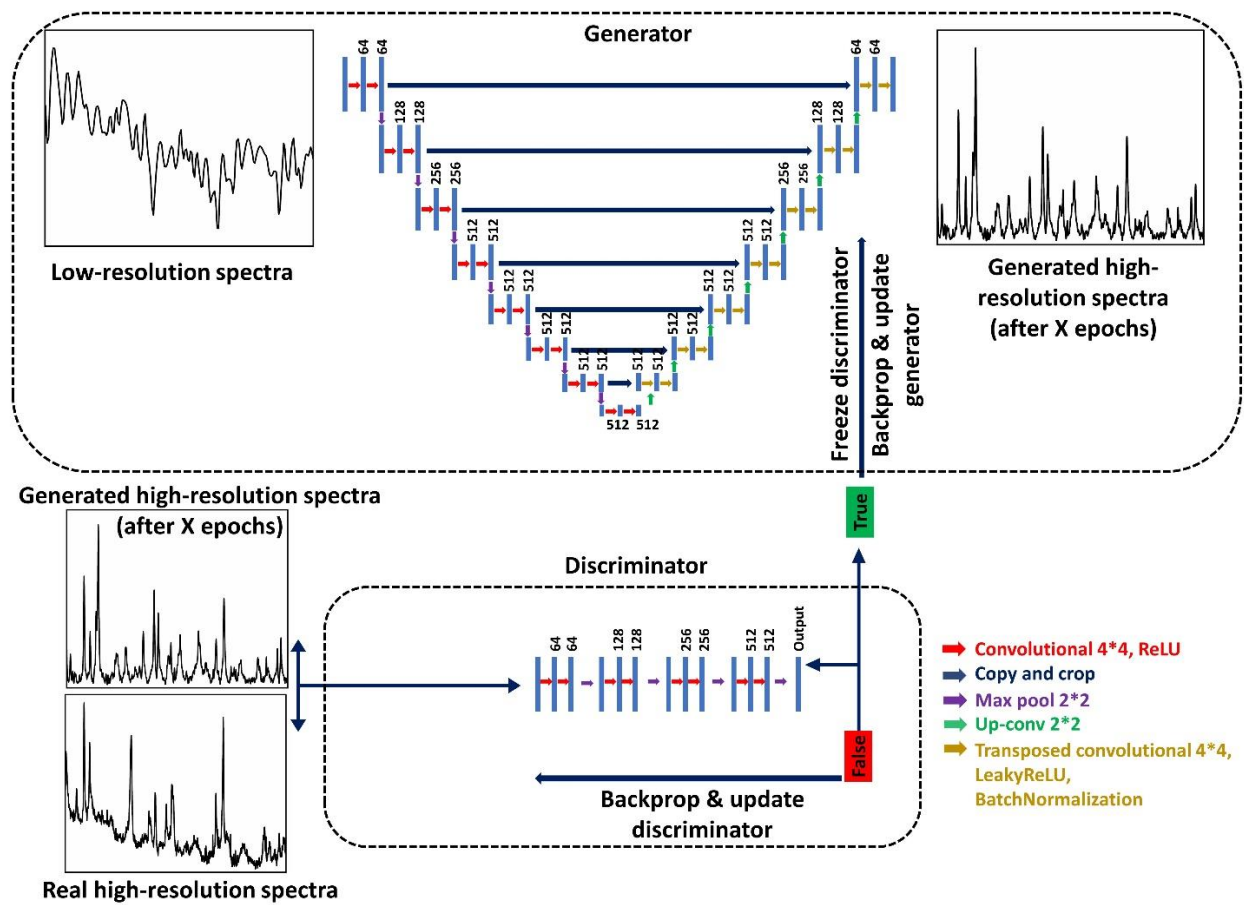

**Figure 3:** The schematic of the Generative Adversarial Network (GAN) model consists of two primary components: a generator and a discriminator. The generator is tasked with producing high-resolution spectra from low-resolution inputs, aiming to generate outputs that closely resemble realistic high-resolution spectra. Conversely, the discriminator's role is to differentiate between the output generated by the generator and real high-resolution spectra. The low-resolution Raman spectra is feed to the generator, after every epoch the generated high-resolution Raman spectra was compared with the real high-resolution counterpart by the discriminator. The discriminator tested the capabilities of generator at every epoch, results in the generation of closely resembled high resolution output by the generator after each epoch.

Upon comparing the high-resolution and low-resolution datasets, a stark discrepancy emerges, wherein spectral features are either inadequately expressed or entirely absent, potentially attributed to a lower grating system and lower focal length. As shown in **Figure 2D**, there is clearly a pattern in the noisy spectra, which results in a good classification among different classes. However, the pertinent question arises: can we extract additional spectral features from these low-resolution datasets? Moreover, is leveraging the poor SNR signals to extract meaningful information plausible? These questions suggest that bridging the gap between low-resolution and high-resolution spectra becomes imperative.

## Leveraging GAN for generating high-resolution Raman spectra

The GAN architecture for generating high-resolution Raman spectra comprises a generator and discriminator. In our implementation, the generator adopts an eight-block U-Net-based design featuring both an encoder and a decoder with multiple layers, as detailed in the model construction section. The overall structure and model flow are illustrated in **Figure 3**. The model takes low-resolution spectra as input, which are then fed into the generator component. Subsequently, the generator generates high-resolution outputs compared to the actual high-resolution signals. The disparity between the generated and actual signals is quantified as the Generator Loss (G Loss). Through iterative epochs, the generator aims to approximate the actual high-resolution results. Conversely, the discriminator's role is to discern between the generated and authentic high-resolution signals. As training progresses, the generator attempts to produce high-resolution signals indistinguishable from the real ones, ultimately deceiving the discriminator. The discrepancy in the discriminator's ability to discriminate between real and generated data is denoted as Discriminator Loss (D Loss). Essentially, the generator seeks to generate data that confounds the discriminator, while the discriminator's task is to differentiate between genuine and generated data accurately. **Figure 3** depicts the workflow of the training process and the resulting generated high-resolution spectra. The model exhibited commendable performance during both the training and validation phases. The initial learning rate was set to 0.0001, with a chosen mini-batch size of one. The model underwent iterative refinement with 177 epochs and 424800 total iterations. In the initial stages, the model was trained using six different samples, with approximately 400 Raman spectra collected from each sample. These samples encompassed various pharmaceutical drugs and organic samples, providing a diverse array of spectral fingerprints to evaluate the model's accuracy. The accuracy obtained in the processed Raman spectra confirms the capability of GAN to retain most of the original spectral features from low spectral compositions without introducing artificial interference during the recovery process. While the accuracy of the model could be further enhanced with a larger dataset and additional sample sets, the primary objective of this study could be elucidated, that is, the utilization of low-resolution Raman signals from a handheld spectrophotometer to generate high-resolution signals.

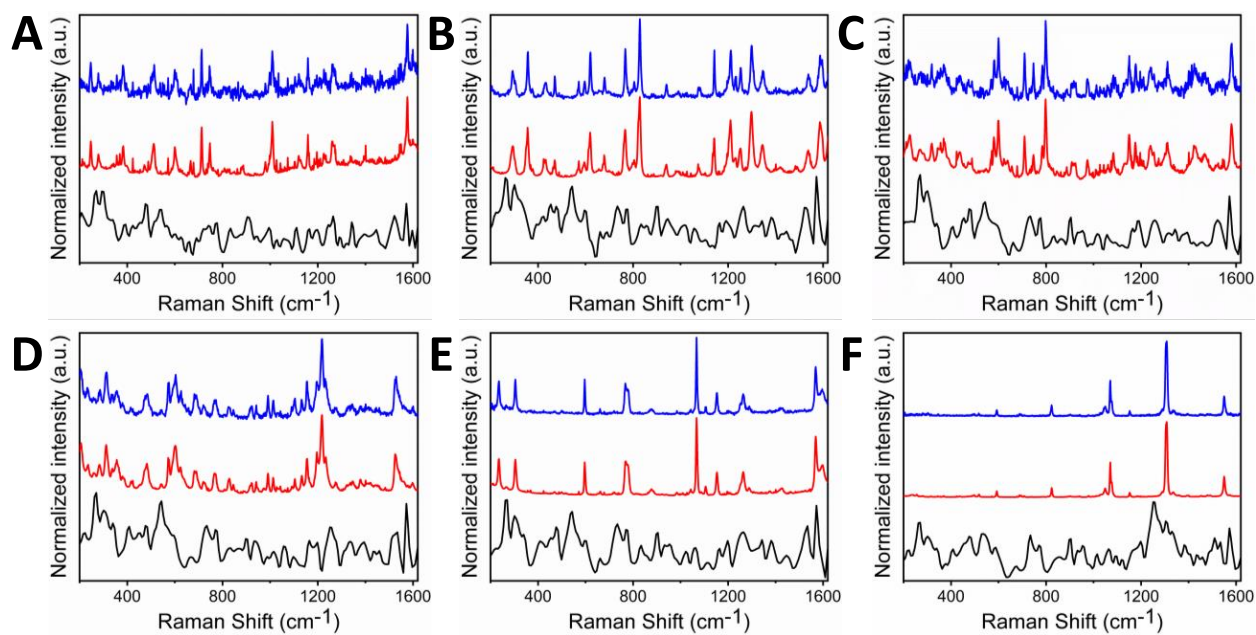

**Figure 4:** Raman spectra from low-resolution (black), high-resolution (red), and GAN generated (blue) of different pharmaceutical and drug molecules, (A) aspirin, (B) ibuprofen, (C) PCM, (D) ranitidine, (E) 4-MBA, and (F) 4-NTP. 400 spectra were collected from each sample.

The loss function equation for the generator and discriminator is given by,

$$G\ Loss = -mean(\log(Y_{generated}))$$

$$D\ Loss = -mean(\log(Y_{real})) - mean(\log(1 - Y_{generated}))$$

Where, $Y_{generated}$ is the generated Raman spectra at a particular epoch and $Y_{real}$ is the real high-resolution Raman spectra. Once the model is fully trained, then the trained model produces significantly realistic high-resolution Raman spectra outputs corresponding to low-resolution inputs, as depicted in **Figure 4**. Consequently, we have a comprehensive dataset comprising both generated high-resolution and actual high-resolution data across all sample classes. The subsequent step entails constructing a classification model to differentiate between these sample classes and developing a method that can take low-resolution inputs from handheld spectrometers and output the corresponding class of the sample along with its generated high-resolution Raman spectra. To address this classification task, we opt for an ANN model consisting of an input layer, some hidden layers, and a classification layer at the end. The detailed network architecture setup was mentioned in the material and method section. The ANN architecture, depicted in **Figure S1**, accommodates real high-resolution datasets introduced into the network through an input layer. The input dataset was partitioned into three subsets—70% for training, 15% for testing, and 15% for validation—to assess model accuracy rigorously. The training was extended over some epochs through iteration, resulting in an ANN model accuracy of approximately 100%. Subsequently, confusion plots and ROC plots were generated for further analysis and illustrated in **Figure S2** and **S3**. The confusion matrix shows how the true class of the sample differs from that of the predicted class of that sample. Since the ANN model was trained using real high-resolution data, the accuracy is almost 100%, as depicted by the confusion matrix in **Figure S2**. Once the model is trained on real high-resolution data, we check the model accuracy over the additional dataset, which is the GAN-generated high-resolution data. The GAN-generated high-resolution data was never fed to the network during the training state. Before the data input, normalization was performed to ensure uniform processing within the network. The model shows more than 88% accuracy for the classification of generated high-resolution data, which is significant and can further be improved, as shown in **Figure S4**. Therefore, by integrating the two networks mentioned above, i.e., GAN and ANN, we can accurately determine the true class of a sample and visualize its high-resolution spectra. The SNR for the low, high, and generated high-resolution Raman spectra were calculated by the below formula:[46]

$$SNR = \frac{Raman\ Peak\ Intensity}{Standard\ deviation\ of\ background\ noise}$$

$$SNR\ (dB) = 10\log_{10}(SNR)$$

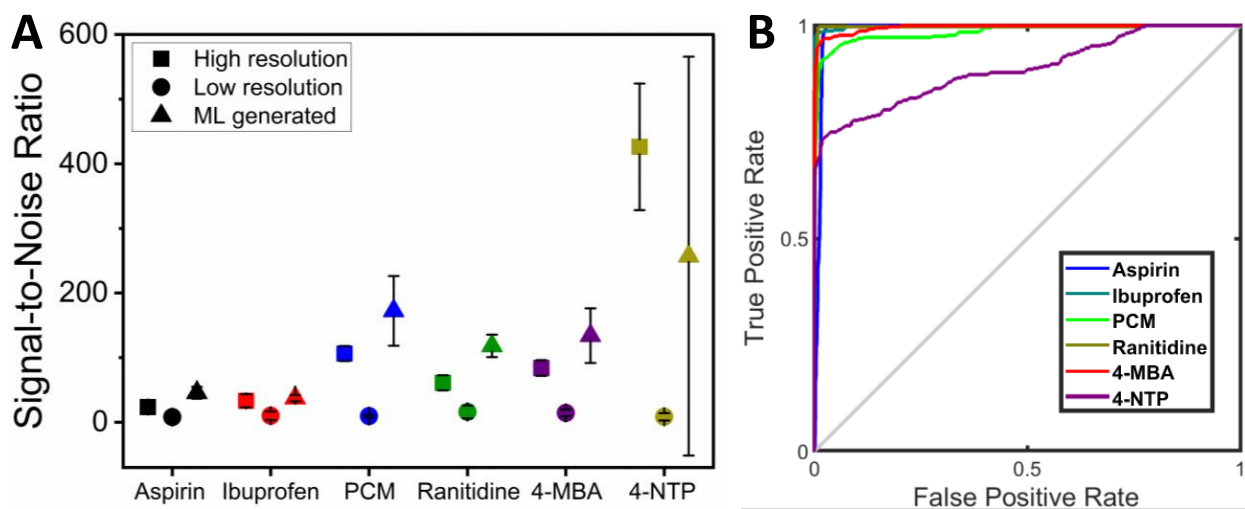

**Figure 5:** (A) SNR calculation of different pharmaceutical and organic molecules using high-resolution spectrometer (square), low-resolution spectrometer (circle), and GAN generated (triangle) results. (B) ROC curve for the classification ANN model for different samples. The closer the curve to the one, more accurate will be the classification.

The calculated SNR of the Raman signal acquired using low-resolution and high-resolution spectrometers is shown in **Figure 5A**. The SNR was determined for the peak with the highest magnitude in the Raman spectrum, i.e., in the case of the 4-NTP molecule corresponding to the symmetric ring breathing vibration of the nitro group at 1346 cm$^{-1}$. Additionally, the standard deviation of the background noise, encompassing all spectral fluctuations of the noise, was computed. The SNR was then calculated by dividing the peak intensity by the standard deviation of the background noise. Notably, the average SNR of the low-resolution Raman spectra typically ranged from 7 to 15, whereas for the high-resolution Raman spectra, it spanned from 25 to 430, depicting a substantial enhancement in the case of the generated spectra. The SNR observed in the GAN-generated high-resolution Raman spectra sometimes surpasses that of real high-resolution Raman spectra, as shown in **Table S2**. This enhancement is primarily attributed to the ML model's capability to discern Raman features while filtering out background noise. **Figure 5B** also shows the ROC (Receiver Operating Characteristic) curve for classification of the low-noise and generated high-resolution data showing a high level of classification. Essentially, the model reconstructs spectral features from noisy low-resolution spectra, inadvertently amplifying the SNR of the Raman spectra during this process. Consequently, the GAN not only transforms low-resolution Raman spectra into high-resolution counterparts but also augments the SNR, particularly benefiting samples characterized by weak scattering. Notably, the proposed GAN model offers several distinct advantages: Firstly, the entire transformation process is typically very fast once optimized. Secondly, parameter adjustments are unnecessary as all parameters are predetermined during the training phase, enabling robust handling of various noise types within the training dataset. Thirdly, it exhibits superior and reproducible capabilities, which is particularly beneficial for Raman spectra with low SNRs.

## Identification of pharmaceutical compounds through spectral barcoding

Several studies have explored the application of Raman barcoding for various purposes, such as detecting counterfeit drugs and developing SERS nanosensors.[8,33] These studies typically utilize commercial Raman spectrometers to acquire high-resolution Raman spectra for subsequent barcoding analysis.[47] In our approach, we have extended the concept of Raman barcoding to low-resolution spectra obtained using portable spectrometers. A key innovation lies in the integration of GAN, which rapidly enhances the spectral features of low-resolution spectra, enabling the generation of fully resolved outputs suitable for barcoding. Consequently, the resulting barcode contains sufficient spectral information for accurate sample identification.

For the barcoding process, we employed the "findpeaks" function in MATLAB, utilizing a peak prominence threshold set at 0.3, indicating that peaks with an SNR of 0.3 or higher are considered valid peaks within the Raman spectrum, discernible from the background noise. Once all peaks within the Raman spectrum and their FWHM values were identified and labeled, an in-house MATLAB algorithm was developed to fetch this information and integrate it into constructing the Raman barcode. The process of making and workflow of the Raman barcode was illustrated in

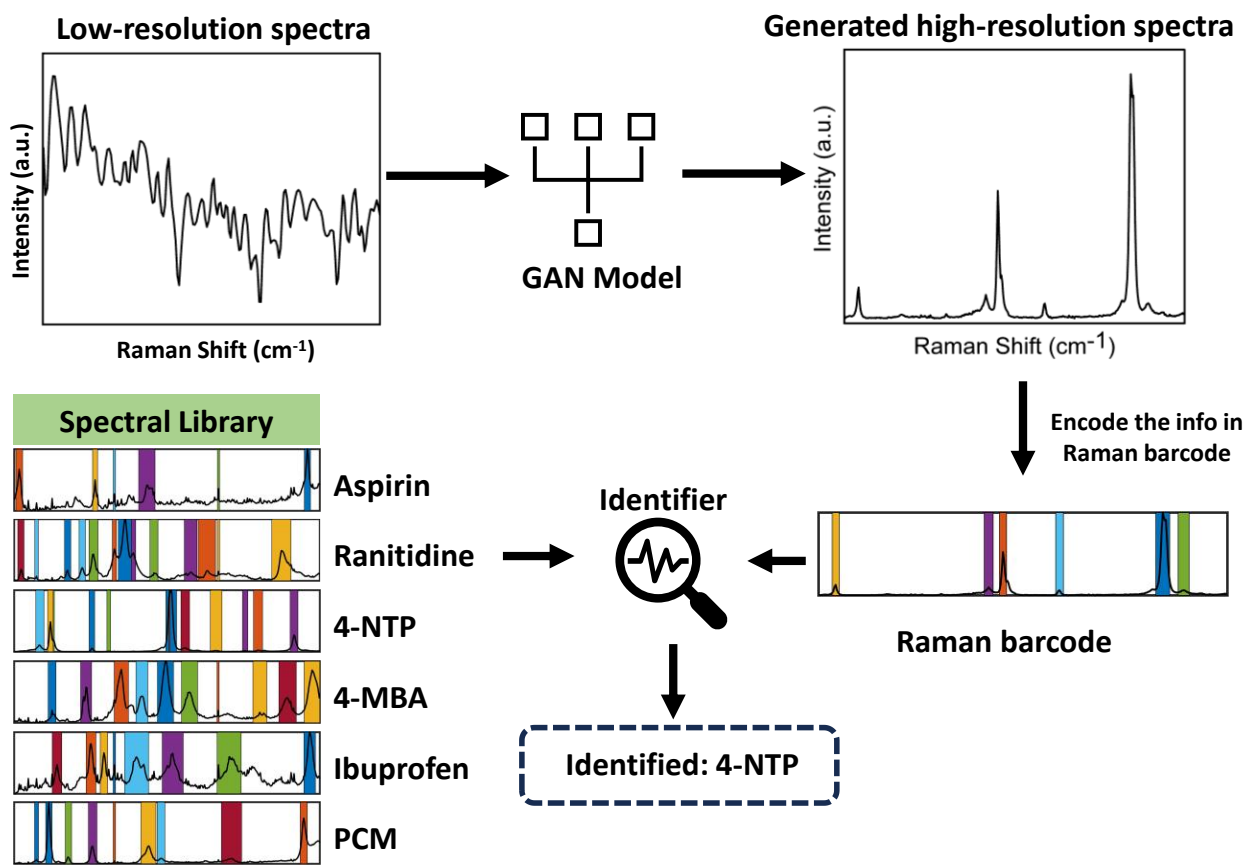

**Figure 6:** The generated high resolution Raman spectra from low resolution through GAN model was converted into Raman barcode. These barcodes were employed as inputs to the model, which performed spectral comparison against a spectral library. The model output provided identification of the unknown sample based on spectral matching.

**Figure S5** and **Figure 6**. Since Raman is a very sensitive technique for the detection of a sample, and every sample has a distinct Raman fingerprint region, the Raman barcode will be a unique identity that is assigned to the sample on the basis of Raman spectra. Consequently, each barcode encapsulates specific information unique to each sample, facilitating streamlined identification of unknown samples. We have created a spectral barcode library using high-resolution Raman spectra obtained from a high-resolution spectrometer. Given that the GAN has successfully transformed low-resolution Raman spectra into their high-resolution equivalents. The identifier then uses these generated high-resolution spectra to extract the FWHM and peak positions, converting them into Raman barcodes. These barcodes are subsequently scanned against the Raman spectral barcode library, and the resulting matrix represents the percentage similarity index. We use the built-in MATLAB function 'SSIM' to calculate the similarity index of the barcode from an unknown sample against the spectral library. The highest similarity index from the barcode library will identify the unknown sample. The barcode-based model effectively identified the compounds from the GAN-generated spectra at an accuracy of nearly 100%. Hence, the implementation of Raman barcode versatility ensures comprehensive sample identification beyond predefined categories, enhancing the utility and effectiveness of the classification model.

## Conclusion

This study presents a methodology for transforming low-resolution input signals obtained from a cost-effective handheld spectrometer into their high-resolution equivalents using a robust GAN model. This setup enables instantaneous classification results that resemble those derived from high-resolution data. We demonstrate the efficacy of a GAN-based ML model for cost-effective and real-time decision-making setups, particularly for the in-depth analysis of pharmaceutical drugs and organic samples. The sample selection process was randomized to ensure a comprehensive assessment of Raman spectral features across the fingerprint region. Low-resolution and high-resolution Raman spectra were acquired using a small handheld spectrophotometer and a fully equipped spectrophotometer of higher resolution. We successfully bridged the resolution gap by integrating the GAN algorithm, accurately mimicking high-resolution counterparts from low-resolution inputs. This effective collaboration between ML-based models and Raman spectroscopy reduces costs and facilitates on-site detection. This approach opens new avenues for miniaturization and the development of home-built spectrophotometers, enabling practical analysis and diagnosis of unknown samples with enhanced accuracy.

## Author Contributions

VY performed the experiments and data analytics using machine learning. VY, SS and AKT were involved in conceptualizing, analysis, data interpretation and writing of the manuscript.

## Funding Source

This work was supported by the IIT Delhi SEED grant (IITD/Plg/ Budget/2020-2021/204758), CSIR Research Grant (01(3042)/21/ EMR-II), and DST SERB (SRG/2020/000440).

## Acknowledgement

This work was supported by the IIT Delhi SEED grant (IITD/Plg/ Budget/2020-2021/204758), CSIR Research Grant (01(3042)/21/ EMR-II), and DST SERB (SRG/2020/000440). VY thanks UGC, Govt. of India, for the student fellowship. The authors thank IIT Delhi HPC facility for computational resources.

## Supporting Information Available

Additional details about the ANN model with its architecture, confusion matrices for accuracy of the trained model, and validation of the model on the additional dataset, i.e., GAN-generated high-resolution Raman data followed by ROC curves. Also, a brief about the barcode-making process is included.

# Machine learning driven high-resolution Raman spectral generation for accurate molecular feature recognition

Vikas Yadav,[‡] Abhay Kumar Tiwari,[⊥] Soumik Siddhanta,[‡*]

[‡]*Department of Chemistry, Indian Institute of Technology Delhi, Hauz Khas, New Delhi, India, 110016*

[⊥]*SPATIALTY.AI, Tagya Research Private Limited, Rajankunte, Bangalore 560064*

*Corresponding author e-mail: soumik@iitd.ac.in

**Table S1:** Raman band assignment of different samples: aspirin, ibuprofen, PCM, ranitidine, 4-MBA, and 4-NTP.[1,2,3,4,5]

| Raman Shift (cm$^{-1}$) | Raman band assignment |
|---|---|
| 1030 | Aromatic rings |
| 1200 | -OH group substitution |
| 1300 | C-C bond |
| 1600 | C=O stretching |
| 700-830 | γ C-H |
| 1231 | C-C ring stretching |
| 858 | Ring breathing |
| 791 | CNC ring stretching |
| 1319 | Amide III |
| 1558 | Amide II N-H in plane deformation |
| 1607 | skeletal aryl C-C ring stretching |
| 1644 | Amide I |
| 841 | out-of-plane C-H bending |
| 1441, 1472 | aryl C-C stretch |
| 1511 | aryl C-H symmetric bends |
| 1587 | NO$_2$ asymmetric stretching |
| 1554, 1408 | NO$_2$ symmetric stretching |
| 1483, 1376 | C-N symmetric stretching |
| 1450 | C-C symmetric stretching |
| 1306 | C-H in-plane bending |
| 1263, 1248 | C-H out of plane bending |

**Table S2:** SNR of Raman spectra accumulated using high resolution (HR), low resolution (LR), and machine learning generated (ML) for different samples.

| Name | No. of Samples | Mean | Standard Deviation | Median | Minimum | Maximum |
|---|---|---|---|---|---|---|
| Aspirin HR | 400 | 23.58176 | 8.79591 | 23.09082 | 12.61017 | 180.80869 |
| Aspirin LR | 400 | 7.99599 | 3.29201 | 7.6165 | 3.94684 | 57.17247 |
| Aspirin ML | 400 | 45.64161 | 9.07993 | 44.74725 | 33.64713 | 162.21038 |
| Ibuprofen HR | 400 | 33.21321 | 10.27409 | 31.84045 | 10.13911 | 177.38332 |
| Ibuprofen LR | 400 | 10.15884 | 6.66026 | 8.20264 | 3.80088 | 59.72808 |
| Ibuprofen ML | 400 | 37.40251 | 4.80574 | 37.17291 | 10.50112 | 55.47692 |
| PCM HR | 400 | 106.31269 | 11.58307 | 106.82571 | 20.51726 | 129.52435 |
| PCM LR | 400 | 9.46906 | 2.24131 | 9.14041 | 4.89948 | 23.58947 |
| PCM ML | 400 | 172.40985 | 54.07065 | 184.89031 | 27.76558 | 314.10354 |
| Ranitidine HR | 400 | 61.02186 | 11.66157 | 60.63593 | 22.27034 | 104.75477 |
| Ranitidine LR | 400 | 15.96444 | 9.97238 | 13.59722 | 5.4066 | 115.82591 |
| Ranitidine ML | 400 | 118.22499 | 17.32099 | 117.96073 | 41.31493 | 187.57732 |
| 4-MBA HR | 400 | 83.94236 | 12.21468 | 84.01384 | 38.64827 | 126.68066 |
| 4-MBA LR | 400 | 14.46693 | 5.20116 | 13.52151 | 6.17104 | 39.74992 |
| 4-MBA ML | 400 | 133.95862 | 42.24702 | 143.86413 | 21.52214 | 232.14881 |
| 4-NTP HR | 400 | 426.2633 | 97.90562 | 409.24547 | 169.19863 | 668.27542 |
| 4-NTP LR | 400 | 8.5804 | 5.70461 | 5.46376 | 3.57232 | 37.4824 |
| 4-NTP ML | 400 | 257.09559 | 308.59492 | 27.9709 | 9.77899 | 1067.8729 |

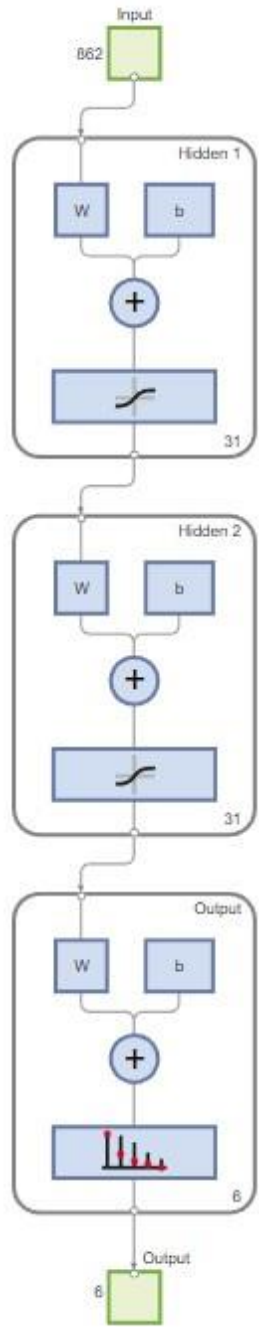

**Figure S1:** The architecture of the ANN classification network consists of one input layer, two hidden layers with 31 neurons each, followed by the classification layer in the end.

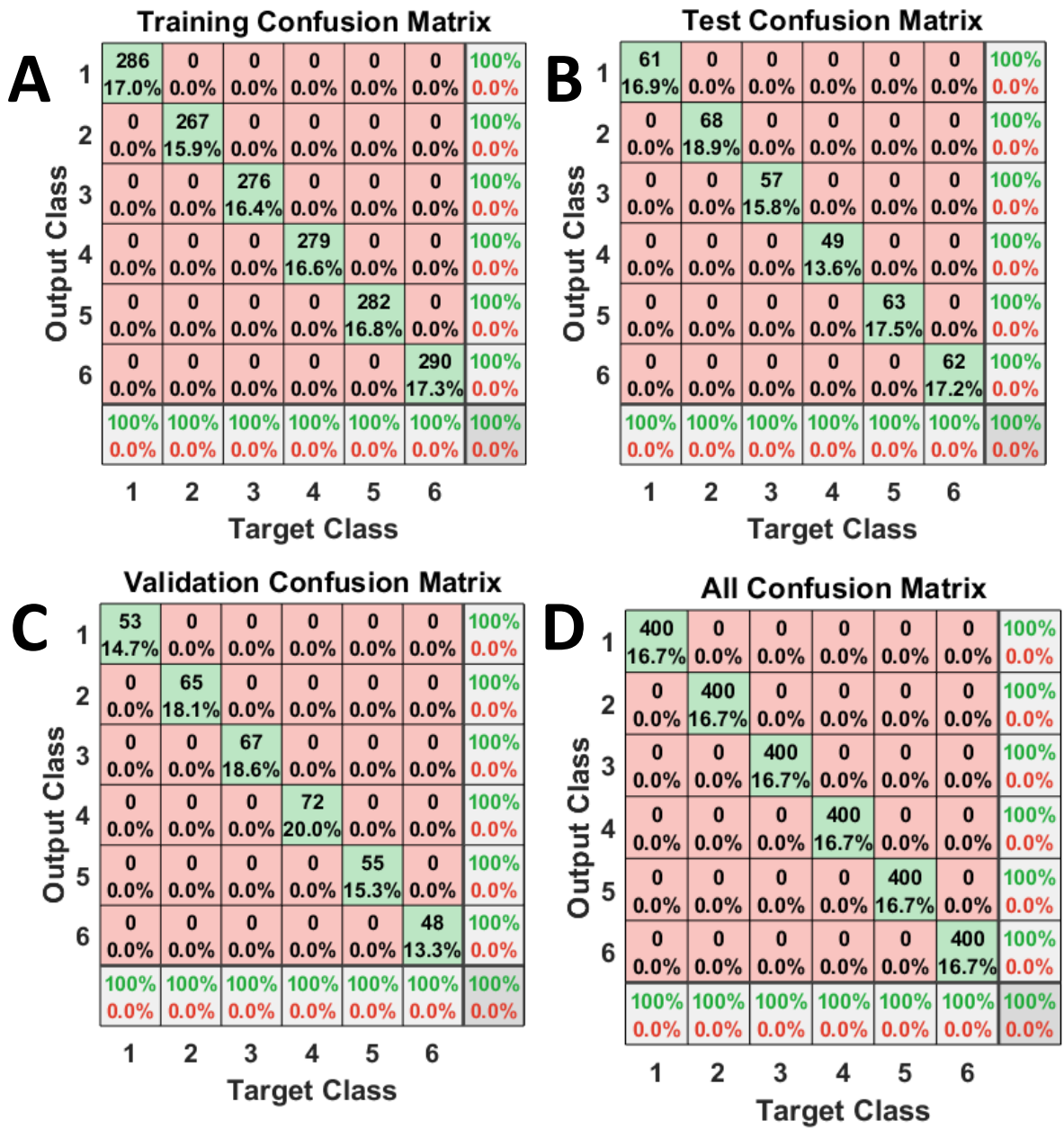

**Figure S2:** Confusion matrices showing the accuracy of the ANN classification network for (A) training, (B) testing, (C) validation, and (D) overall accuracy. The representation for different samples is as follows: (1) aspirin, (2) ibuprofen, (3) PCM, (4) ranitidine, (5) 4-MBA, and (6) 4-NTP. The confusion matrix shows the target class (actual) vs output class (predicted by the model).

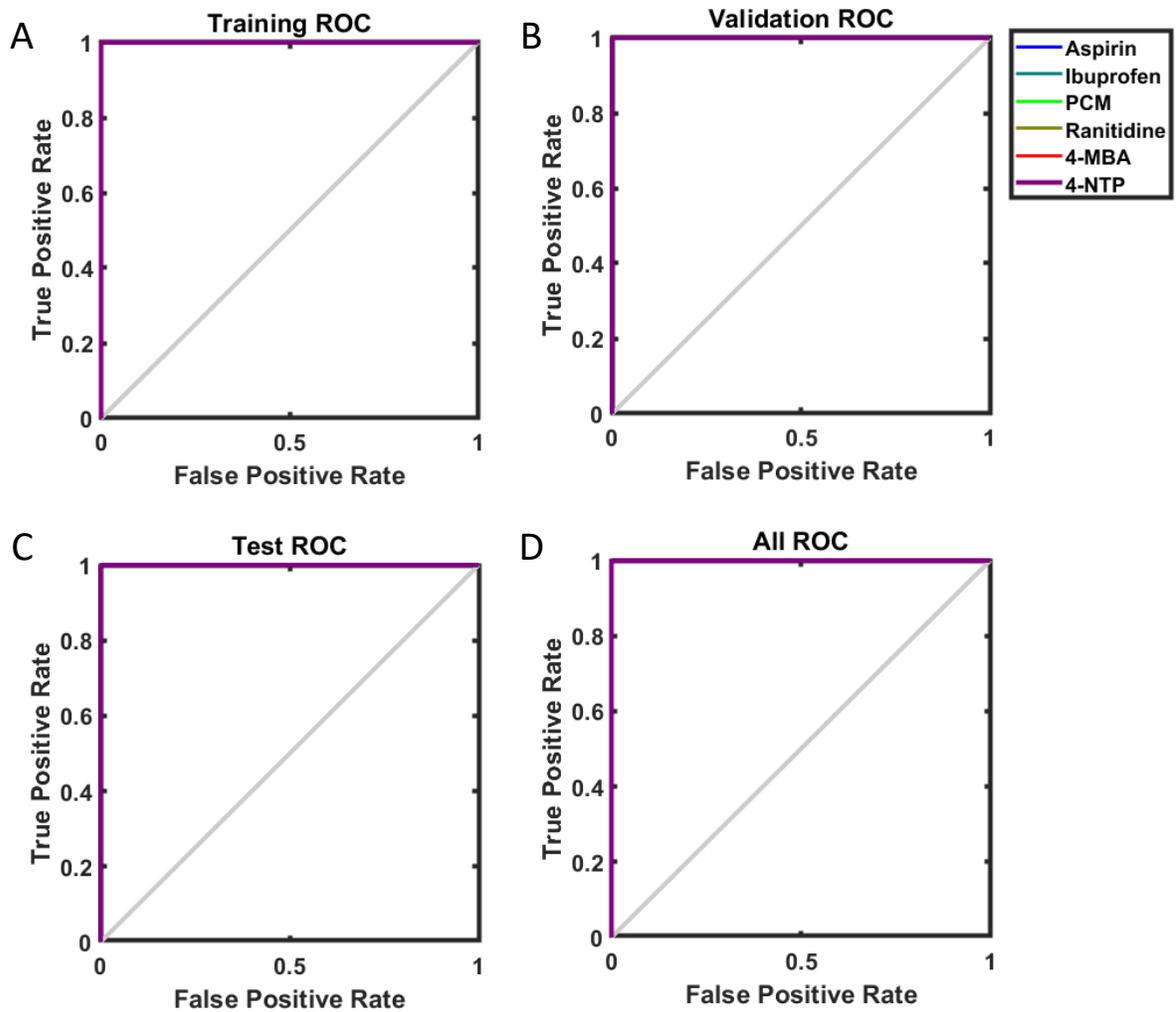

**Figure S3:** ROC of the ANN classification network for (A) training, (B) validation, (C) testing, and (D) overall. The closer the curve is to the one, the more accurate the results will be for classification.

|   | 1 | 2 | 3 | 4 | 5 | 6 | |
|---|---|---|---|---|---|---|---|
| **1** | **399** 16.6% | 0 0.0% | 0 0.0% | 1 0.0% | 2 0.1% | 58 2.4% | 86.7% 13.3% |
| **2** | 0 0.0% | **392** 16.3% | 0 0.0% | 0 0.0% | 0 0.0% | 10 0.4% | 97.5% 2.5% |
| **3** | 0 0.0% | 1 0.0% | **382** 15.9% | 0 0.0% | 0 0.0% | 127 5.3% | 74.9% 25.1% |
| **4** | 0 0.0% | 7 0.3% | 18 0.8% | **399** 16.6% | 32 1.3% | 2 0.1% | 87.1% 12.9% |
| **5** | 0 0.0% | 0 0.0% | 0 0.0% | 0 0.0% | **366** 15.2% | 8 0.3% | 97.9% 2.1% |
| **6** | 1 0.0% | 0 0.0% | 0 0.0% | 0 0.0% | 0 0.0% | **195** 8.1% | 99.5% 0.5% |
|   | 99.8% 0.2% | 98.0% 2.0% | 95.5% 4.5% | 99.8% 0.2% | 91.5% 8.5% | 48.8% 51.2% | **88.9%** **11.1%** |

(Output Class vs. Target Class)

**Figure S4:** Confusion matrices showing the accuracy of the ANN classification network for the additional dataset, i.e., GAN generated high-resolution Raman data.

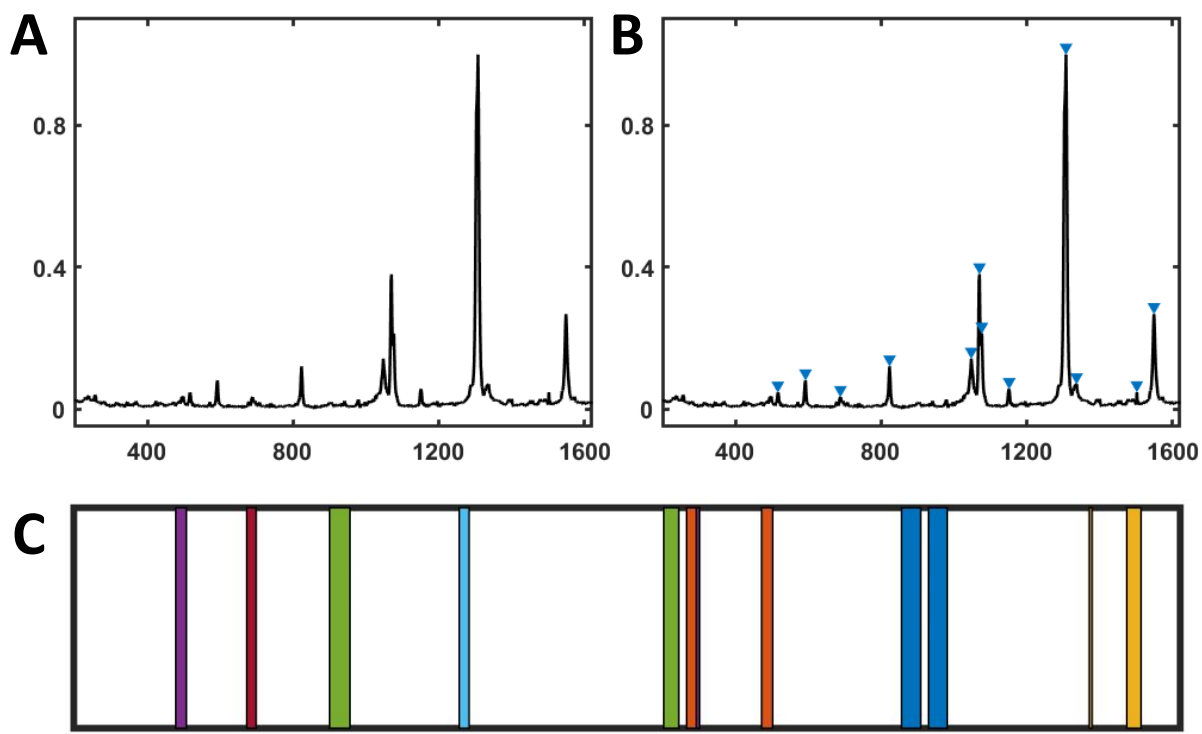

**Figure S5:** The process of making a Raman barcode. (A) the Raman spectra of the 4-NTP molecule. (B) using the findpeaks function to extract the peak location and FWHM and incorporate that information in the barcode. (C) The Raman barcode of the 4-NTP molecule.